\documentclass[letterpaper,amsmath,
  amssymb,aps,prd,nofootinbib,
singlecolumn,
superscriptaddress,altaffilsymbol,12pt]{revtex4-1}

\usepackage{amsmath}
\usepackage{amsfonts}
\usepackage{amssymb}
\usepackage{graphicx}
\usepackage{psfrag}
\usepackage[hang]{footmisc}
\usepackage{setspace}
\newcommand{\beq}{\begin{equation}}
\newcommand{\eeq}{\end{equation}}
\newcommand{\bea}{\begin{eqnarray}}
\newcommand{\eea}{\end{eqnarray}}
\newcommand{\beqn}{\begin{equation*}}
\newcommand{\eeqn}{\end{equation*}}
\newcommand{\bean}{\begin{eqnarray*}}
\newcommand{\eean}{\end{eqnarray*}}

\newcommand*{\cref}[1]{Chapter~\ref{#1}}

\begin{document}
\title{Evolution of the universe during the inflationary epoch}

\author{Gabriel Germ\'an}
\affiliation{Instituto de Ciencias F\'{\i}sicas, 
Universidad Nacional
Aut\'onoma de M\'exico, Cuernavaca, Morelos, 62210, Mexico}

\date{\today}
\begin{abstract}
We often find in the literature solutions to the Friedmann and fluid equations for simple cosmological models during the matter, radiation or cosmological constant dominated epochs. However no solutions appear for the inflationary era dominated by the potential energy of a scalar field due, perhaps, to the fact that we do not have as yet a strongly favored model of inflation; there are, of course, very well motivated models which fit the data. The purpose of this article is to study with some detail the evolution of the Universe during inflation in the slow-roll approximation. Taking the Starobinsky model as an example, we display exact solutions for the time evolution of the scalar field $\phi(t)$, scale factor $a(t)$, Hubble function $H(t)$, equation of state parameter $\omega(t)$ and acceleration of the scale factor $\ddot{a}(t)$ among other quantities of interest.
\end{abstract}
\maketitle
\section{Introduction}
When studying cosmological models it is typical to find simple examples in the literature where the energy density of the universe is dominated by radiation, matter and often, when trying to describe inflation \cite{Guth:1980zm}-\cite{Martin:2018ycu}, by a cosmological constant. It is also possible to find variations of the above by studying mixtures where several of these energy components appear simultaneously in the Friedmann equation. Although we do not have yet a definitive model of inflation, we do have several very interesting and well-motivated models  \cite{Martin:2013nzq}, \cite{Martin:2013tda} that could be used as examples instead of the cosmological constant model, with more realistic results.

With this in mind we study the Friedmann and fluid equations in the slow-roll (SR) approximation and reach general equations from where exact analytical solutions can be obtained. In particular, we have carried out a detailed study of the Starobinsky model \cite{Starobinsky:1980te}-\cite{Whitt:1984pd} where we depart from the standard procedure  and obtain exact analytical solutions of the dynamical equations. We obtain solutions in the SR approximation for the time evolution of quantities such as the inflaton field, scale factor and acceleration of the scale factor. Analytical expressions are also given for the equation of state  parameter (EoS), the Hubble function and other quantities of interest. This exercise is briefly extended when writing some of the above quantities as functions of the inflaton field. We work in Planck units where $M_{pl}=2.44\times 10^{18}GeV=1$.

We begin by establishing in general terms the equations that are subsequently solved for the Starobinsky model. The equations  we write below in the SR approximation can be applied to any model of inflation during the inflationary evolution. The fluid equation for an expanding Universe described by a FRW metric is given by 
\beq
\label{EQ}
\ddot{\phi}+3H\dot{\phi}+V^{\prime}=0,
\eeq
where a dot means derivative with respect to time and a prime derivative w.r.t. the inflaton field $\phi$. In the slow-roll (SR) approximation Eq.~(\ref{EQ}) reduces to
\beq
\label{EQsr1}
3H\dot{\phi}+V^{\prime}=0,
\eeq
where, in the SR approximation, $H= \sqrt{V/3}$. From here it follows that
\beq
\label{EQsr2}
\frac{d\phi}{dt}+\frac{2}{\sqrt{3}}\frac{d\sqrt{V}}{d\phi}=0.
\eeq
Thus, the equation from where the solution $\phi(t)$ follows is 
\beq
\label{solfi}
\int\left(\frac{d\sqrt{V}}{d\phi}\right)^{-1}d\phi=-\frac{2}{\sqrt{3}}\,t+cte,
\eeq
with $cte$ a constant here determined by the condition $\phi(0)=\phi_e$ where $\phi_e$ denotes the end of inflation. Thus, the origin of time is chosen as the beginning of the reheating epoch. 

To obtain the time evolution of the scale factor  we use the Friedmann equation in the SR approximation $H= \sqrt{V/3}$ or
\beq
\label{EQda}
\frac{da}{a}=\sqrt{\frac{V(t)}{3}}\,dt,
\eeq 
from where we get
\beq
\label{EQa}
a(t)=a_e\, e^{\int \sqrt{\frac{V(t)}{3}}\,dt},
\eeq
with $a_e$ a constant not fixed by conditions on $a(t)$ or its derivatives. The constant $a_e$ is a model dependent quantity given by $a_e=\frac{k_p}{H_k}e^{N_{ke}}$, where $k_p$ is the pivot scale wavenumber mode, $H_k$ the Hubble function at the beginning of the last $N_{ke}\equiv\ln(\frac{a_e}{a_k})$ e-folds from the time the scale factor was $a_k\equiv a(t_k)$ to the end of inflation at $a_e$.
\section{The Starobinsky inflationary model} \label{STA} 
\begin{figure}[tb]
\begin{center}
\includegraphics[width=8cm]{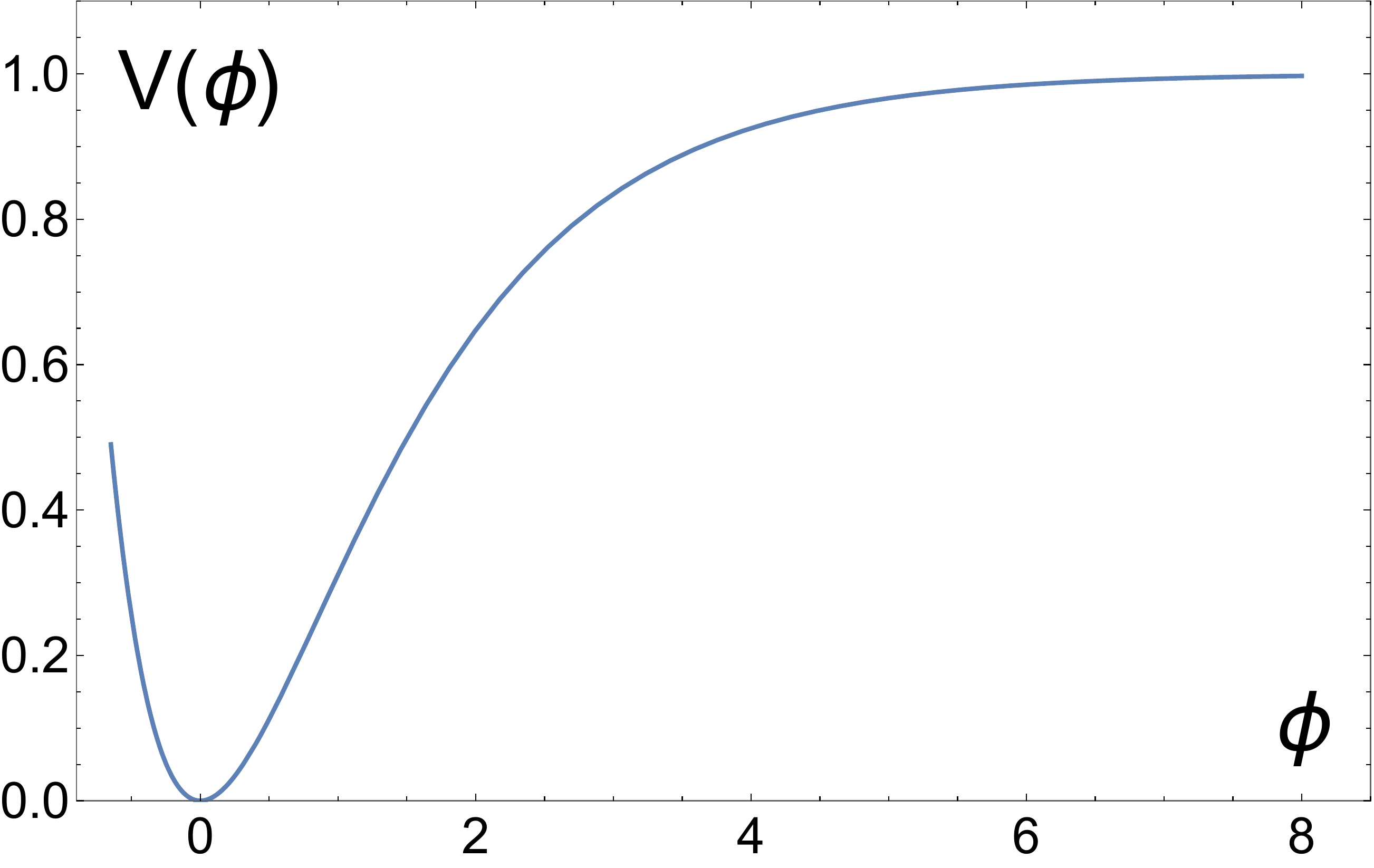}
\caption{\small Schematic plot of the potential given by  Eq.~(\ref{pot}). Scales the size of the pivot scale leave the horizon during inflation at $\phi_k$ somewhere between $\phi=5.21$ to 5.50 (see Eq.~(\ref{bounds})). Inflation ends at $\phi=\phi_e\approx 0.94$ as given by Eq.~(\ref{fie}).}
\label{pot1}
\end{center}
\end{figure}

As a particular example to illustrate the procedure outlined above we study the Starobinsky \cite{Starobinsky:1980te}-\cite{Whitt:1984pd} model of inflation, see also \cite{Bezrukov:2007ep}-\cite{GarciaBellido:2011de} for a Higgs-like model ending in the same potential of Eq.~(\ref{pot}) below. In any case the potential is given by 
\beq
\label{pot}
V= V_0 \left(1- e^{-\sqrt{\frac{2}{3}}\phi} \right)^2,
\eeq
where $V_0$ is an overall constant fixed by the scalar power spectrum amplitude. The potential given by Eq.~(\ref{pot}) is schematically shown in Fig.~\ref{pot1}. 
\subsection{The standard approach} \label{STAs} 
Typically the inflationary epoch is studied by giving the scalar spectral index $n_s$ and tensor-to-scalar ratio $r$ in terms of the number of e-folds during inflation $N_{ke}=-\int_{\phi_k}^{\phi_e}\frac{V}{V^{\prime}}d\phi$ or, using bounds for the spectral index, determine bounds for $r$ as well as for $N_{ke}$ \cite{Cook:2015vqa}.  In this case the solution for $\phi_k$\footnote{The subindex $k$ denotes the value of the inflaton when scales the size of the pivot scale leave the horizon.} in terms of the spectral index $n_s=1+2\eta-6\epsilon$ is
\begin{figure}[tb]
\begin{center}
\includegraphics[width=8cm]{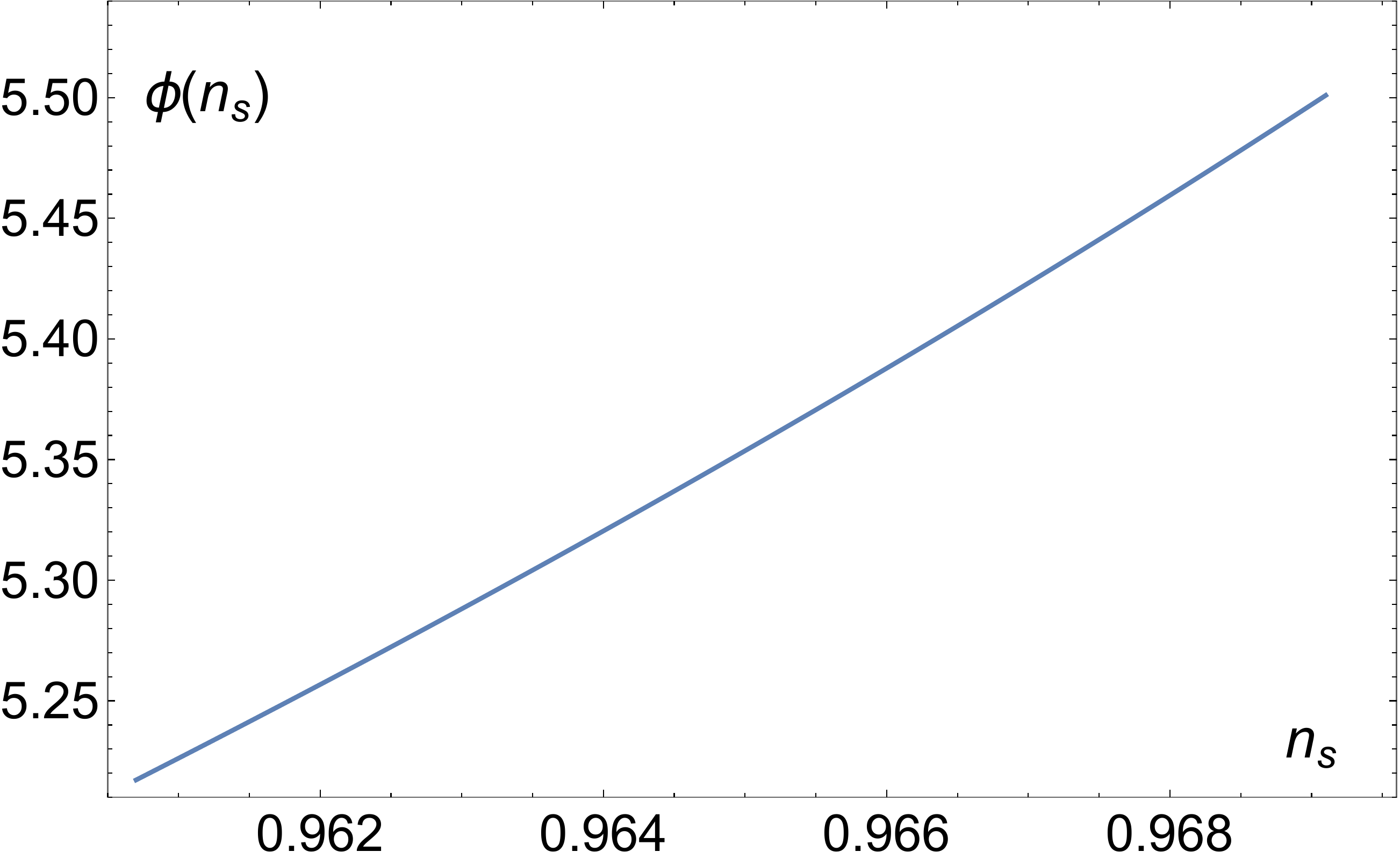}
\caption{\small The solution given by Eq.~(\ref{fik}) is shown for the range of values reported by the Planck Collaboration for the spectral index $n_s=0.9649\pm 0.0042$ \cite{Aghanim:2018eyx}, \cite{Akrami:2018odb}.}
\label{fins1}
\end{center}
\end{figure}
\beq
\label{fik}
\phi_k= \sqrt{\frac {3}{2}}\ln      \left(\frac{7-3n_s+4\sqrt{4-3n_s}}{3(1-n_s)}  \right).
\eeq
This solution is illustrated in Fig.~\ref{fins1}. Using Planck's reported range for the spectral index $n_s=0.9649\pm 0.0042$ \cite{Aghanim:2018eyx}, \cite{Akrami:2018odb} we obtain $\phi_k=5.35\pm 0.15$,
while the end of inflation is given by the solution to the equation $\epsilon =1$ at $\phi_e$:
\beq
\label{fie}
\phi_e= \sqrt{\frac {3}{2}}\ln\left(1+\frac{2}{\sqrt{3}}\right) .
\eeq
The number of $e$-folds during inflation from $\phi_k$ to $\phi_e$ is
\beq
\label{efolds}
N_{ke}= \frac {3}{4} \left( e^{\sqrt{\frac{2}{3}}\phi_k}-e^{\sqrt{\frac{2}{3}}\phi_e}\right) -\frac{\sqrt{6}}{4}\left(\phi_k-\phi_e \right).
\eeq
From the equation for the amplitude of scalar density perturbations at wave number $k$
\beq
\label{A}
A_s(k) =\frac{1}{24\pi ^{2}} \frac{V_k}{\epsilon _k},
\eeq
we obtain an expression for the potential at $\phi_k$
\beq
\label{Vk}
V_k= 3 H_{k}^2=\frac{3}{2}\pi^2 r A_s\,,
\eeq
where $A_s=2.1\times 10^{-9}$ is the amplitude of scalar density perturbations at wavenumber mode $k$ and $V_k$ defines the inflationary energy scale through $\Lambda\equiv V_k^{1/4}$. The tensor-to-scalar ratio is defined as $r\equiv 16\epsilon_k=8\left(\frac{V^{\prime}}{V}\right)^2$ at $\phi_k$ and can be written as a function of the spectral index as follows
\beq
\label{r}
r= \frac {4}{3} \left(5-3n_s-2\sqrt{4-3n_s} \right) .
\eeq
The energy density at $k$ is given by $\rho_k= V_k$ while at the end of inflation is
\beq
\label{roe}
\rho_e= \frac{3}{2} V_e\,,
\eeq
where $V_e$, the potential at the end of inflation, can be related to $n_s$  by means of the formula
\beq
\label{Ve}
V_e= \frac{\left(1- e^{-\sqrt{\frac{2}{3}}\phi_e} \right)^2        }{\left(1- e^{-\sqrt{\frac{2}{3}}\phi_k} \right)^2    }3H_k^2\,.
\eeq
Plots for $r$ and the number of e-folds $N_{ke}$ as functions of the spectral index $n_s$ are shown in Fig.~\ref{rns,efolds}.
\begin{figure}[t!]
\par
\begin{center}
\includegraphics[trim = 0mm  0mm 1mm 1mm, clip, width=8.cm, height=5.cm]{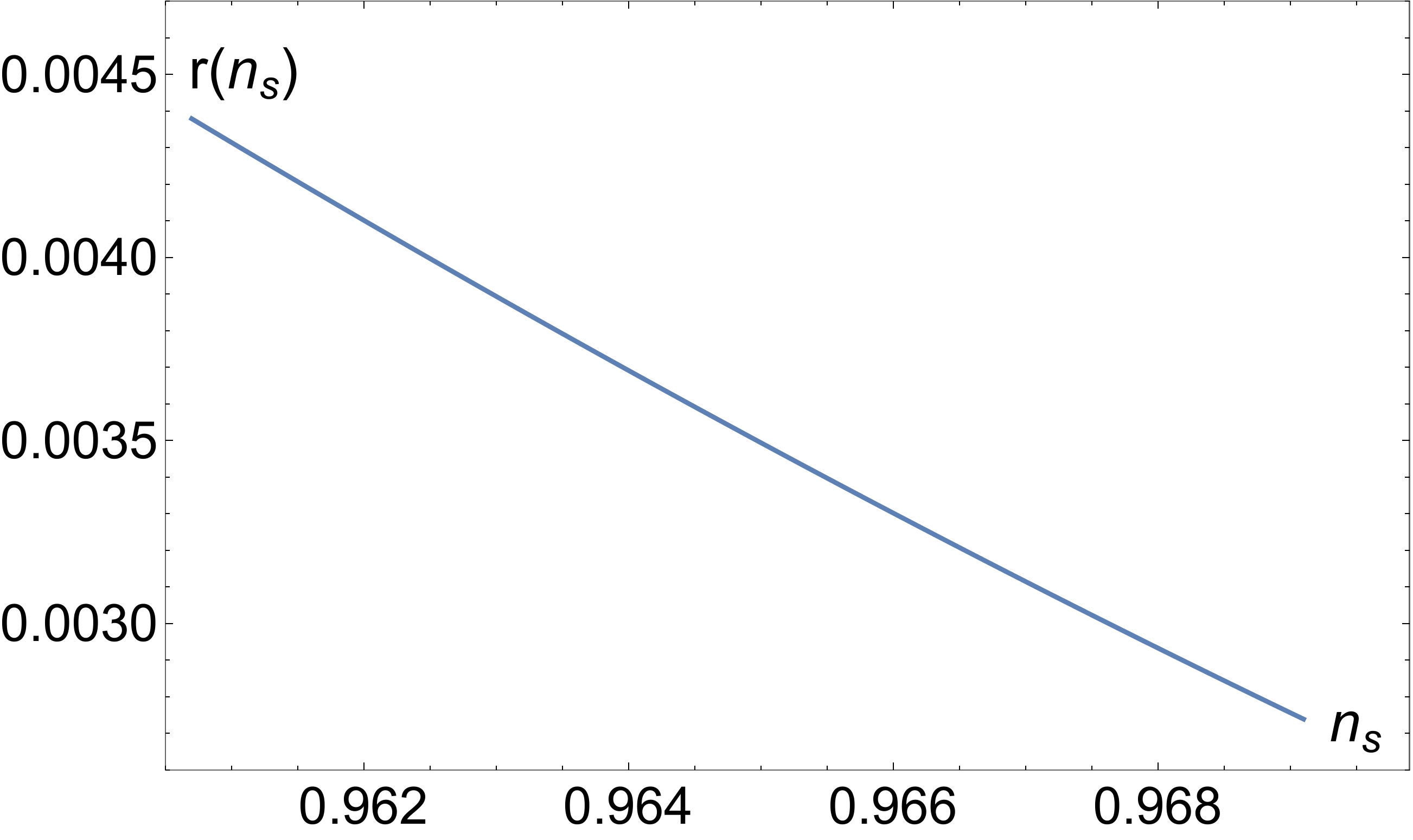}
\includegraphics[trim = 0mm  0mm 1mm 1mm, clip, width=8.cm, height=5.cm]{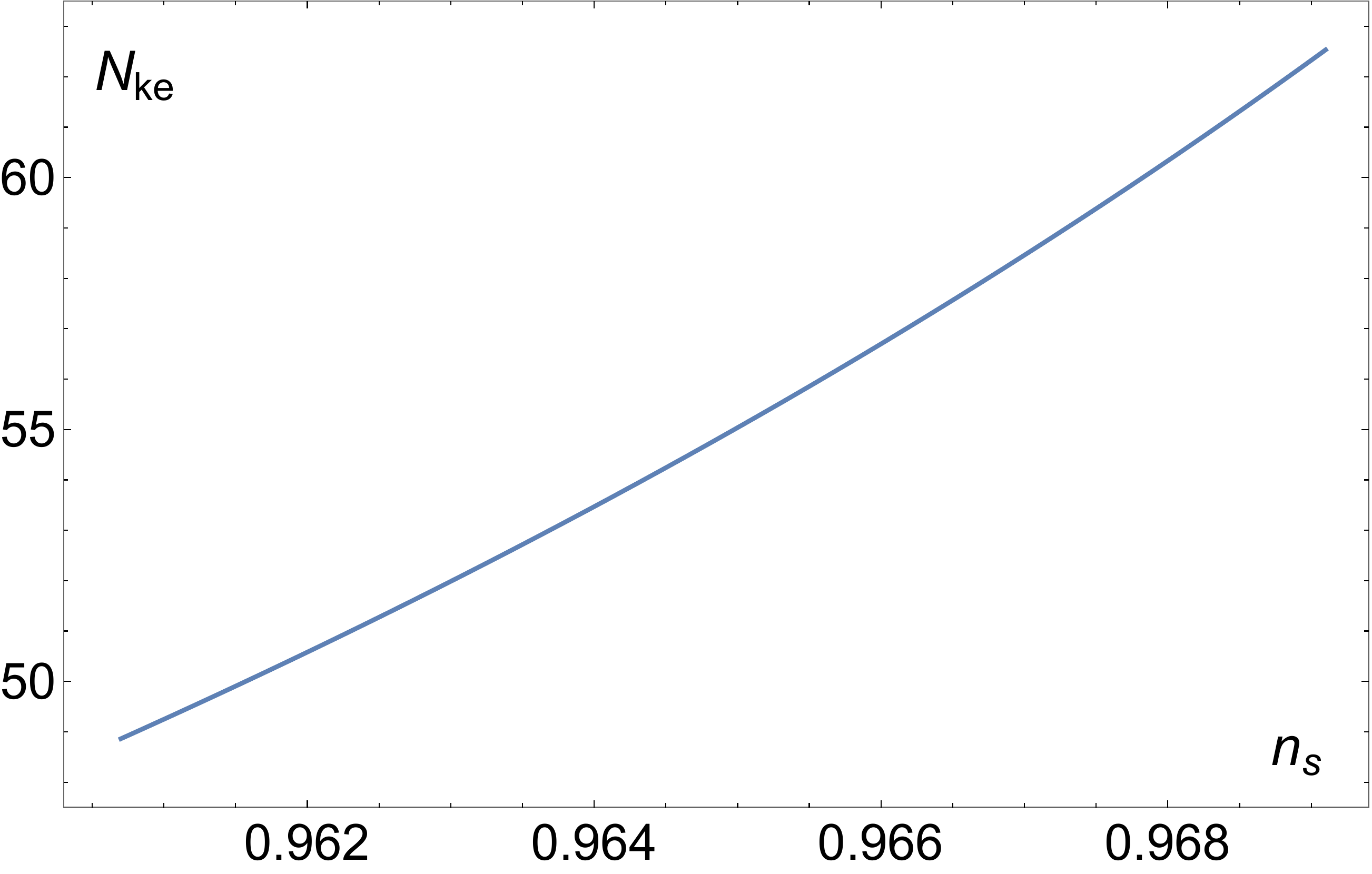}
\end{center}
\caption{Plots of the tensor-to-scalar index $r$, Eq.~(\ref{r}) and the number of e-folds $N_{ke}$, Eq.~(\ref{efolds}) as functions of the spectral index for the range of values reported by the Planck Collaboration, $n_s=0.9649\pm 0.0042$ \cite{Aghanim:2018eyx}, \cite{Akrami:2018odb}.}
\label{rns,efolds}
\end{figure}
\subsection{The time evolution} \label{STAt} 
Solving Eq.~(\ref {solfi}) for the potential (\ref {pot}) gives
\beq
\label{fit}
\phi(t)= \sqrt{\frac {3}{2}} \ln\left(1+\frac{2}{\sqrt{3}}-\frac{4}{3}\sqrt{\frac{V_0}{3}}\,\,t \right) ,
\eeq
an expansion around $t=0$ shows that $\phi(t)=\phi_e+2\left(\sqrt{2}(1-\frac{2}{\sqrt{3}})\sqrt{V_0}\right)t+...$,
which clearly reduces to $\phi_e$ as given by Eq.~(\ref{fie}) for $t=0$. 

From Eqs.~(\ref{fit}) and (\ref{fik}) we get
\beq
\label{tns}
t= \frac{3-2\sqrt{3}-2\sqrt{12-9n_s}-3n_s}{2(1-n_s)\sqrt{V_0}},
\eeq
and from the scalar density perturbation Eq.~(\ref{A})
\beq
\label{V0}
V_0=\frac{2 A_s \pi^2}{\sinh^4(\frac{\phi_k}{\sqrt{6}})},
\eeq
where $\phi_k$ is given by Eq.~(\ref{fik}). From  Eq.~(\ref{V0}) we see that there is a range of values  $V_0$ can take depending on $n_s$ thus, all figures of expressions involving $V_0$ are plotted for the whole range of values quantities can take within the bounds $n_s=0.9649\pm 0.0042$ (shadowed regions in some figures). For this range of $n_s$ values we find that the quantities  $\phi_k$ and $t$ are bounded as follows
\beq
\label{bounds}
0.9607< n_s< 0.9691 \quad \Leftrightarrow \quad 5.21<\phi_k<5.50,  \quad   -7.54\times 10^6>t>-1.21 \times 10^7\,.
\eeq
Thus, the whole of inflation from the time scales the size of the pivot scale left the horizon when the scale factor was $a_k$ to the end of inflation at $a_e$ lasted between (7.54 to 12.1)$\times 10^6$ or putting back time units, between (2.0 to 3.3)$\times 10^{-36}s$.
\begin{figure}[tb]
\begin{center}
\includegraphics[width=8cm]{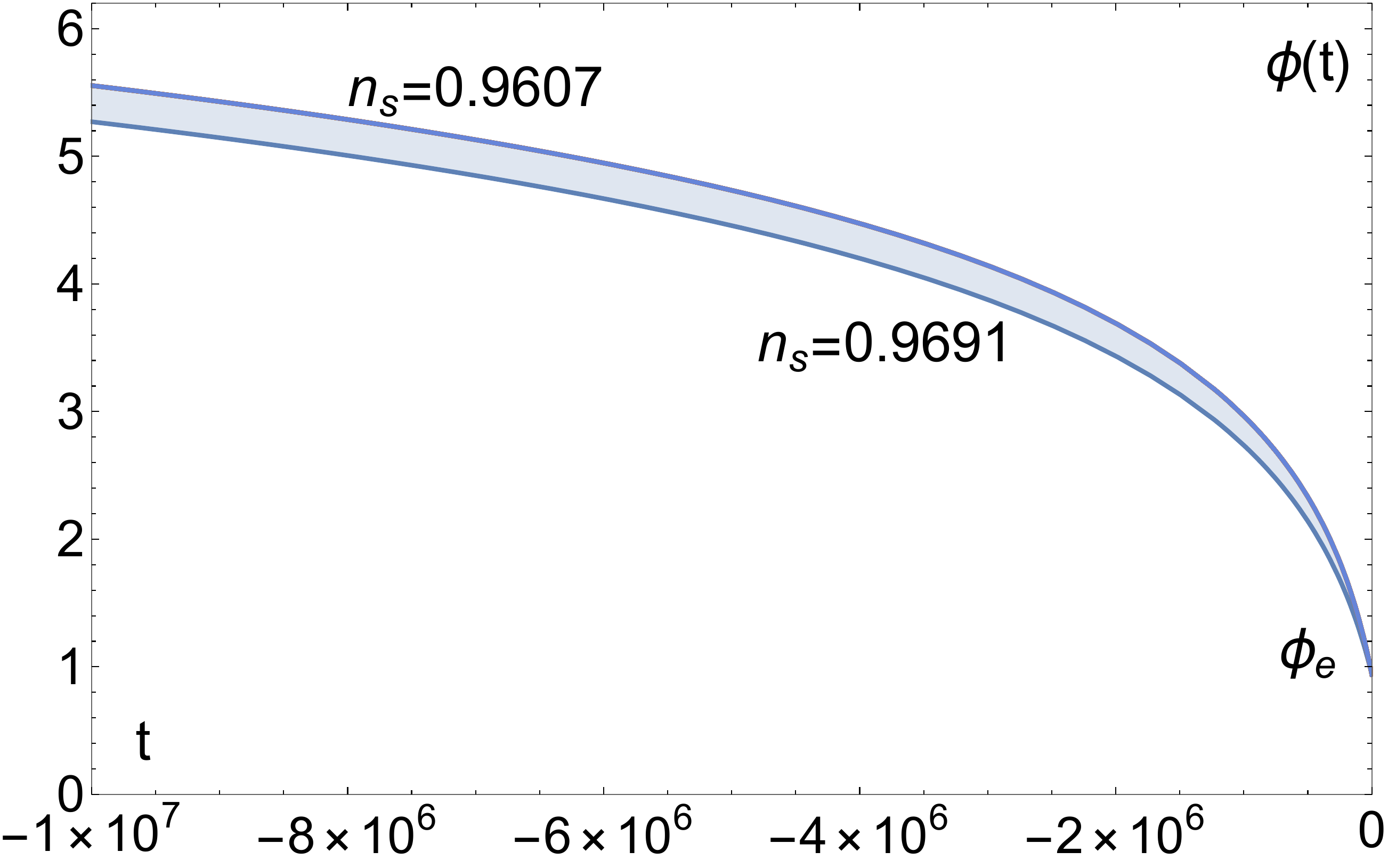}
\caption{\small The shadowed region shows all the possible values $\phi(t)$ can take up to the end of inflation at $\phi_e$ within the Planck range for the spectral index $n_s=0.9649\pm 0.0042$. The two curves corresponding to the bounding values of $n_s=0.9607$ and $n_s=0.9691$ are also shown. }
\label{fit0}
\end{center}
\end{figure}
\begin{figure}[t!]
\par
\begin{center}
\includegraphics[trim = 0mm  0mm 1mm 1mm, clip, width=8.cm, height=5.cm]{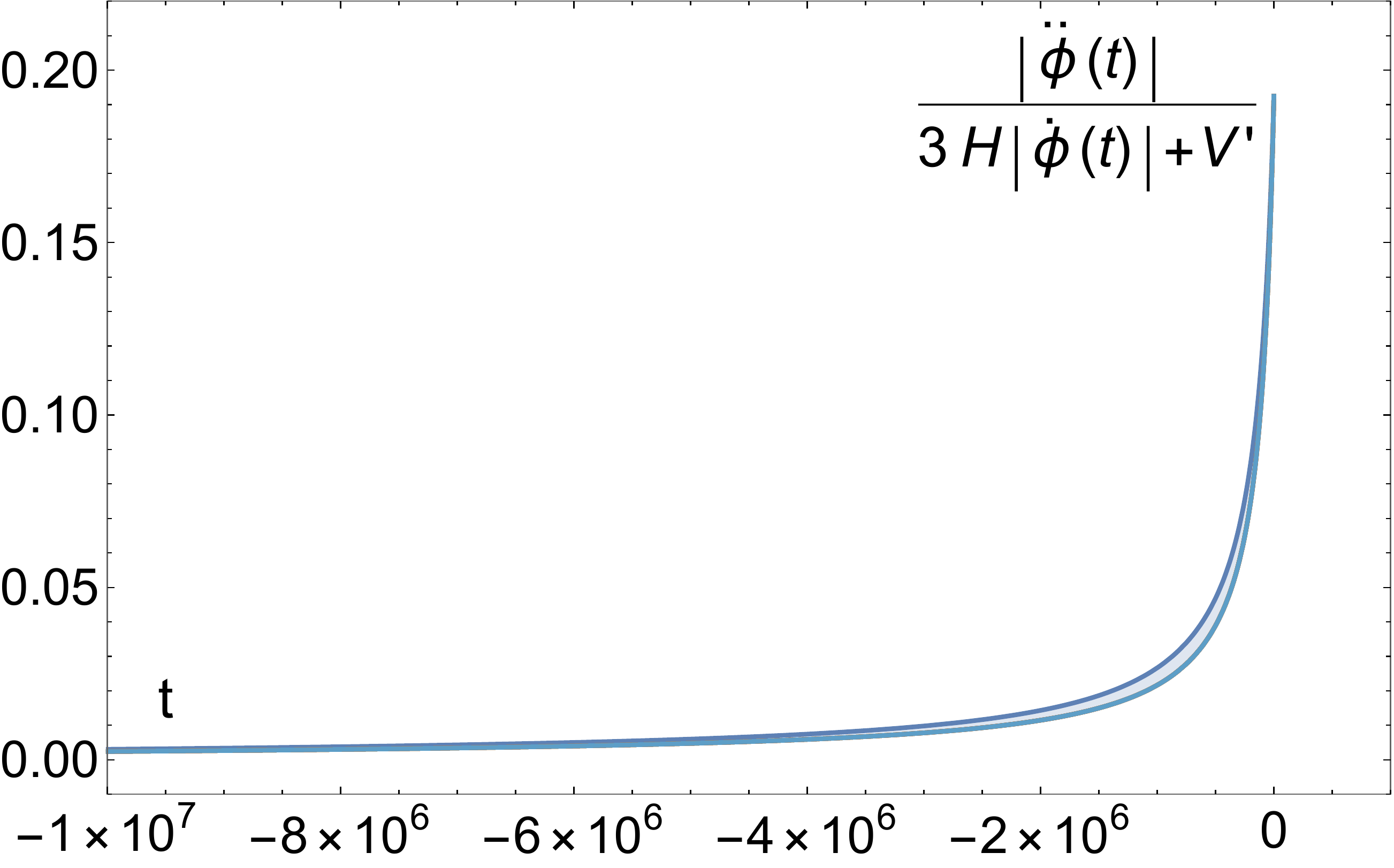}
\includegraphics[trim = 0mm  0mm 1mm 1mm, clip, width=8.cm, height=5.cm]{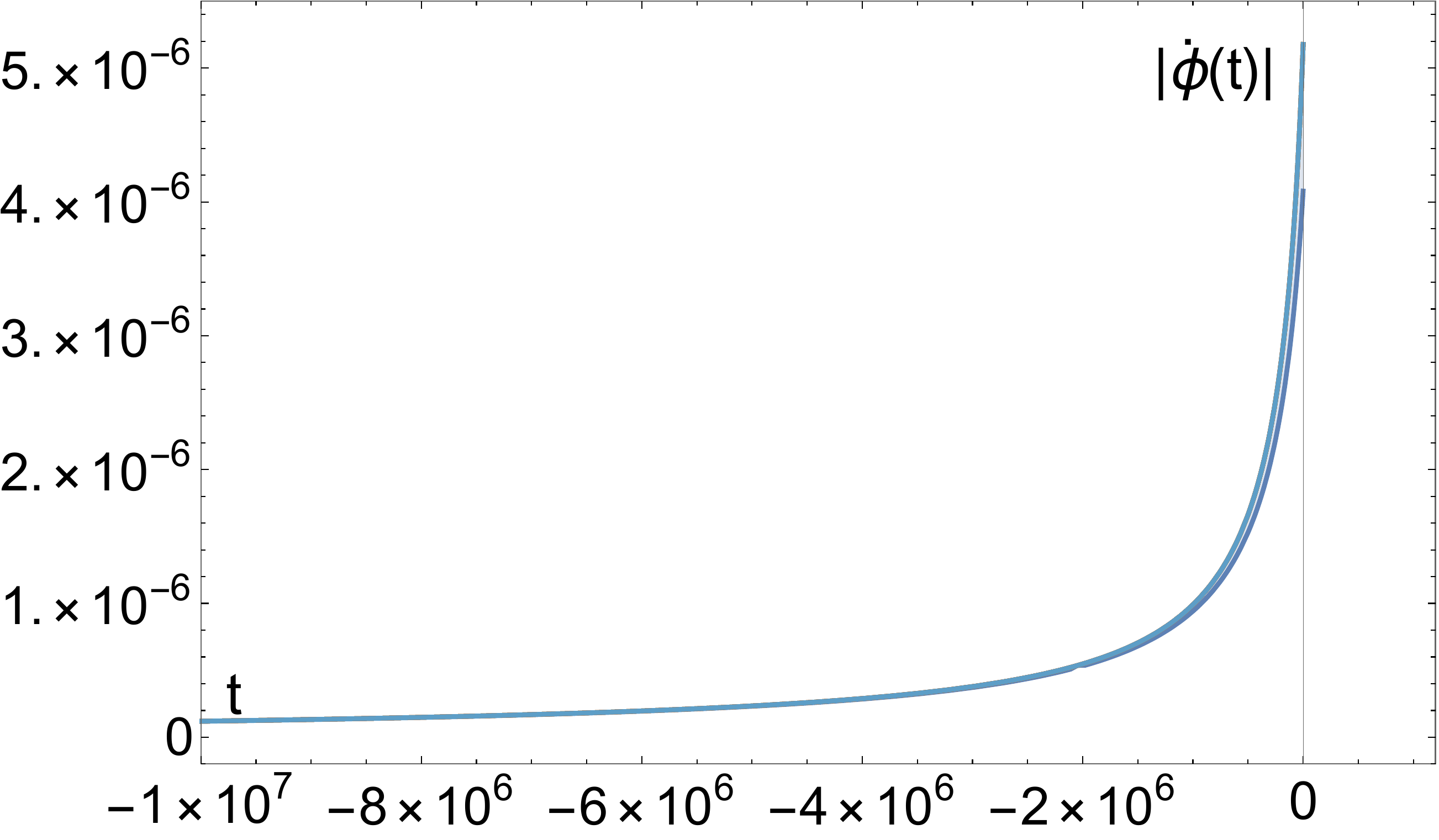}
\end{center}
\caption{In the SR approximation the term  $\ddot{\phi}$ is neglected w.r.t. the term $3H\dot{\phi}+V^{\prime}$. The plot in left hand side panel shows that this assumption is justified for the solution $\phi(t)$ given by Eq.~(\ref {fit}). The term $\ddot{\phi}(t)$ is less than $1\%$ smaller than $3H\dot{\phi}+V^{\prime}$ for most inflation. The right hand panel shows the velocity $\dot{\phi}(t)$ of the inflation slowly rolling during most of inflation. The shadowed region as described in Fig.~\ref{fit0}.
}
\label{SRapprox}
\end{figure}
In Fig.~\ref{SRapprox} we compare the neglected term $\ddot{\phi}$ with $3H\dot{\phi}+V^{\prime}$ where we also show $\dot{\phi}(t)$. We see that most of the time during inflation the term $\ddot{\phi}$ is less that $1\%$ smaller than $3H\dot{\phi}+V^{\prime}$.

\begin{figure}[tb]
\begin{center}
\includegraphics[width=8cm]{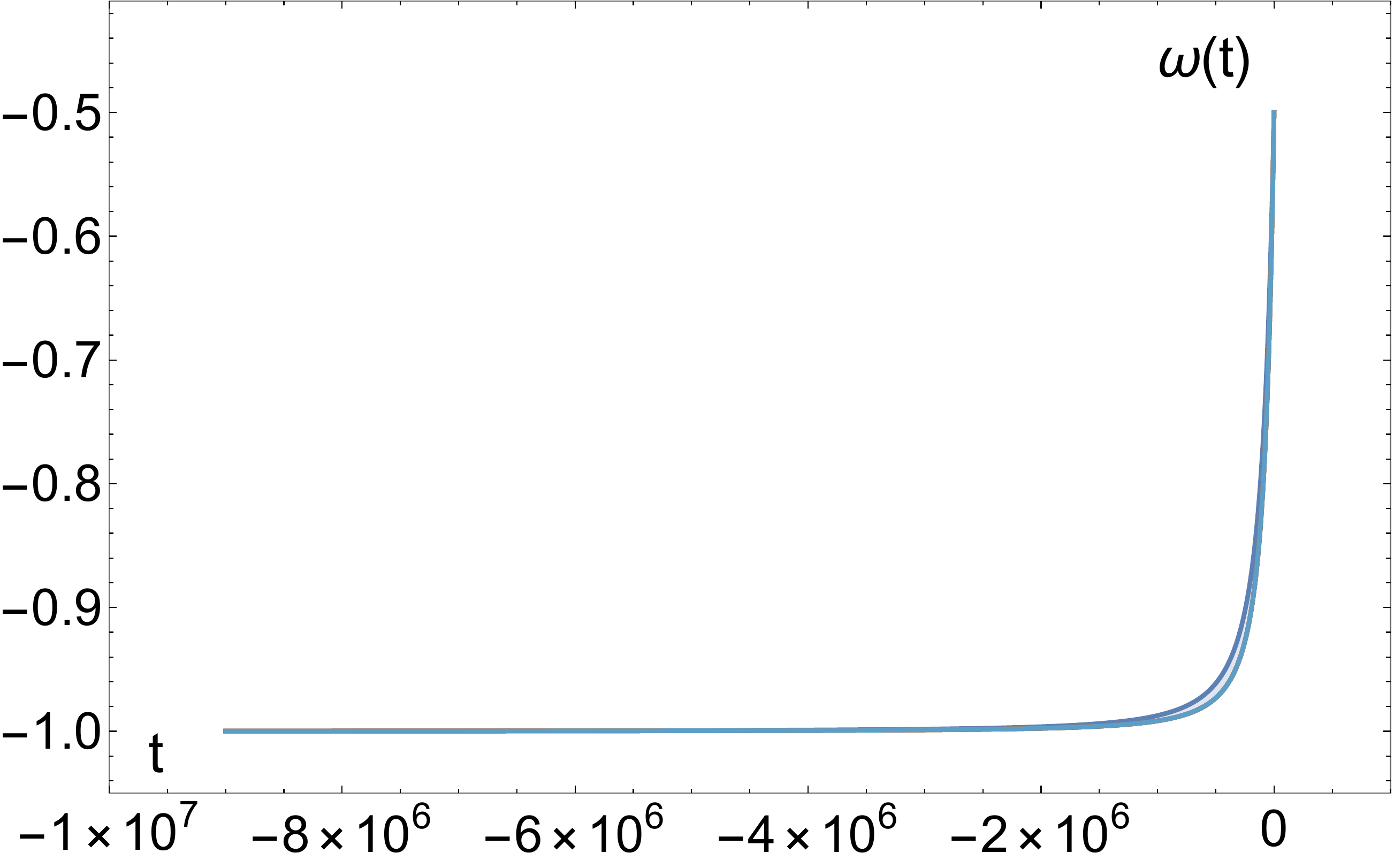}
\caption{\small The time-evolution of the equation of state parameter  Eq.~(\ref{wt}). During most of inflation $\omega(t)$ remains close to -1 increasing just before the end of inflation at $t=0$. The shadowed region as described in Fig.~\ref{fit0}.
}
\label{wt1}
\end{center}
\end{figure}

To write an expression for the equation of state parameter (EoS) we first substitute  Eq.~(\ref {fit}) in (\ref {pot}) to get the time-dependent potential 
\beq
\label{Vt}
V(t)= V_0\left(1-\frac {9}{9+6\sqrt{3}-4\sqrt{3V_0}\,t} \right)^2 ,
\eeq
although in the SR approximation we are neglecting the kinetic energy term in favor of the potential energy we can still write the EoS as follows
\beq
\label{wt}
\omega(t)=\frac {p}{\rho} =\frac{\frac{1}{2}\dot{\phi}^2-V}{\frac{1}{2}\dot{\phi}^2+V}= -1+\frac{3}{6-6\sqrt{V_0}\,t+2V_0\,t^2} ,
\eeq
this equation should be valid whenever $\frac{1}{2}\dot{\phi}^2<<V$. We show $\omega(t)$ in Fig.~\ref{wt1} where we can see that only close to the end of inflation $\omega(t)$ changes appreciably, consistent with our original SR assumption.

\begin{figure}[t!]
\par
\begin{center}
\includegraphics[trim = 0mm  0mm 1mm 1mm, clip, width=8.cm, height=5.cm]{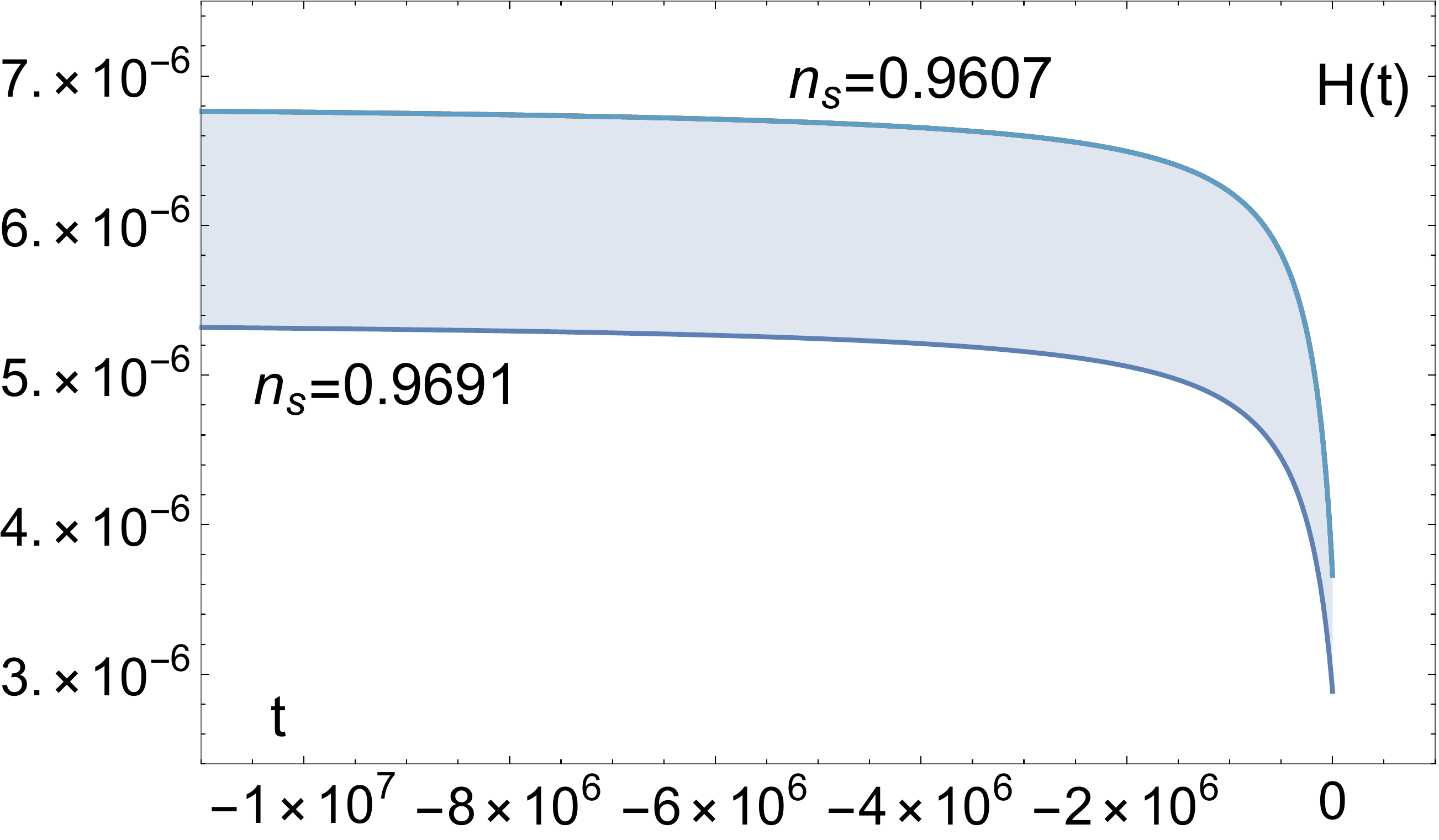}
\end{center}
\caption{The plot shows the evolution of the Hubble function as a function of time. During most of inflation $H(t)$ is almost constant decreasing very slowly and only close to the end of inflation its evolution is appreciable. The shadowed region as described in Fig.~\ref{fit0}.}
\label{Ht1}
\end{figure}

To obtain an expression for the scale factor as a function of time we solve Eq.~(\ref {EQa}) for the potential (\ref {Vt}), the solution is
\beq
\label{at}
a(t)=a_e\,\left(1+\frac{4}{3}(-2+\sqrt{3})\sqrt{V_0}\,t \right)^{3/4}e^{\sqrt{\frac{V_0}{3}}\,t} ,
\eeq
where $a_e$ is the value of $a(t)$ at the end of inflation when $t=0$. We can see the quasi-exponential character of the solution however, close to $t=0$ (where inflation ends),
\beq
\label{atapprox}
a(t)/a_e \approx 1+ (-2+\frac{4}{\sqrt{3}})\sqrt{V_0}\,t +{\cal O}(t^2).
\eeq
\begin{figure}[t!]
\par
\begin{center}
\includegraphics[trim = 0mm  0mm 1mm 1mm, clip, width=8.cm, height=5.cm]{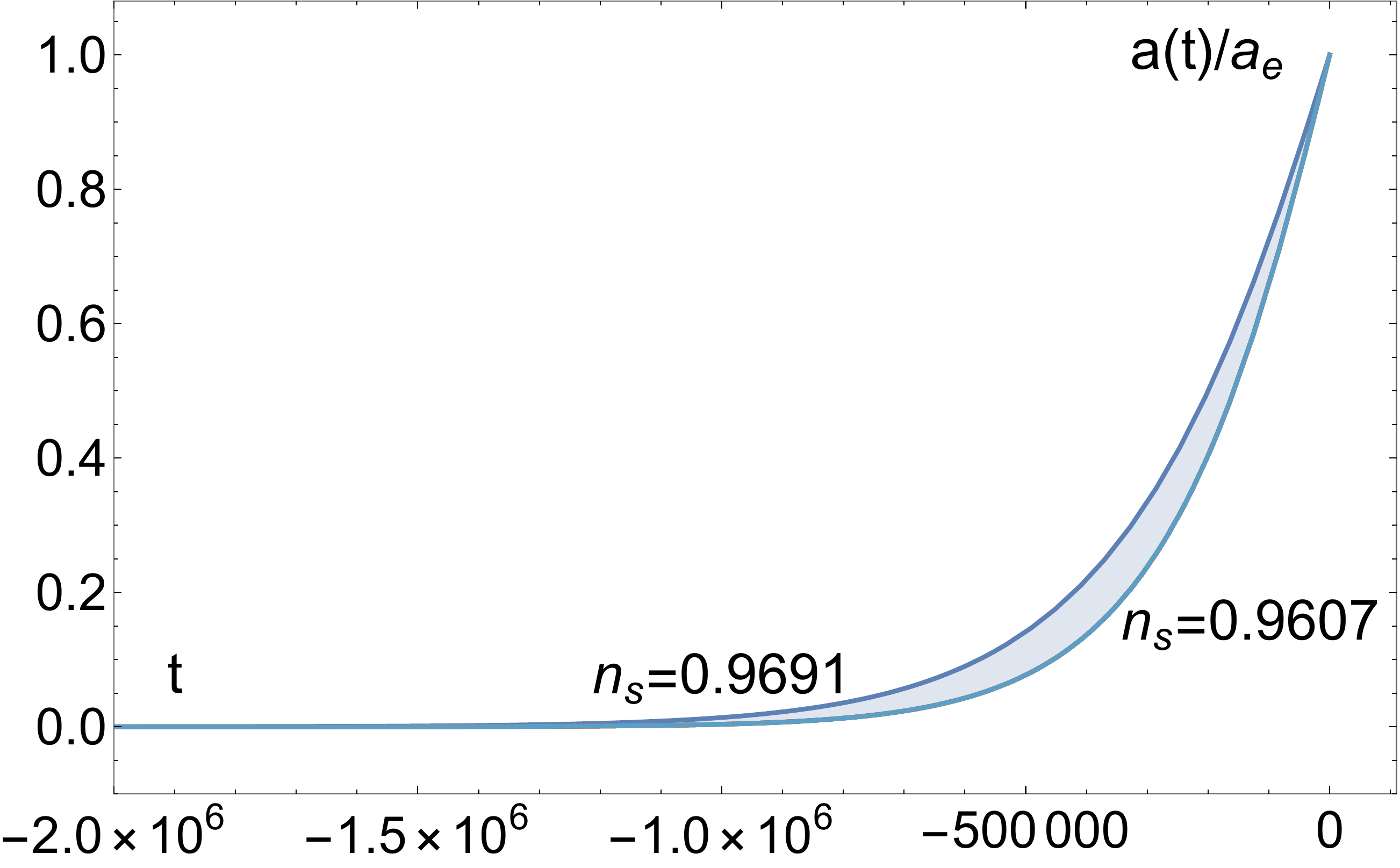}
\includegraphics[trim = 0mm  0mm 1mm 1mm, clip, width=8.cm, height=5.cm]{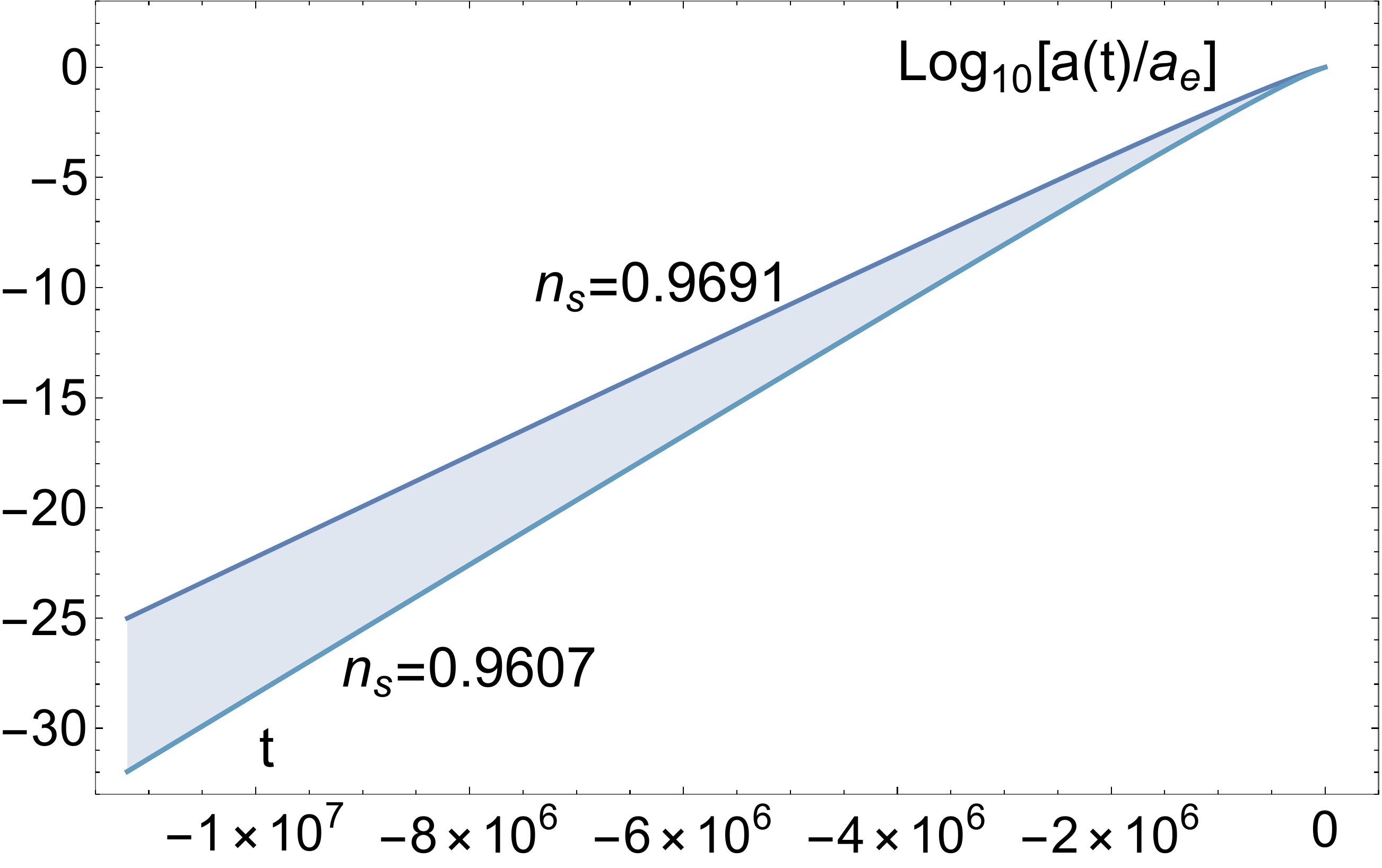}
\end{center}
\caption{The scale factor normalized by its value at the end of inflation is shown as a function of time. The quasi-exponential character is evident from the logarithmic evolution in the r.h.s. panel. For $t=0$ ( the end of inflation) the ratio $a(t)/a_e$ appropriately reaches  the value 1 and  its logarithm consistently reaches zero. The shadowed region as described in Fig.~\ref{fit0}.
}
\label{at1,logat1}
\end{figure}
The time evolution of the wavenumber mode is easily obtained since $k(t)\equiv a(t)H(t)=\dot{a}(t)$, the result is
\beq
\label{k}
k(t)=\frac{(6-4\sqrt{3})(-3+2\sqrt{V_0}\,t)\sqrt{V_0}}{3\times 3^{3/4}(3-4(2-\sqrt{3})\sqrt{V_0}\,t)^{1/4}}a_e\,e^{\sqrt{\frac{V_0}{3}}\,t},
\eeq
while the acceleration of the scale factor $\ddot{a}(t)$ is given by 
\beq
\label{acelt}
\ddot{a}(t)=\frac{16(7-4\sqrt{3})(-3+\sqrt{V_0}\,t)V_0^{3/2}\,t}{3\times 3^{3/4}(3-4(2-\sqrt{3})\sqrt{V_0}\,t)^{5/4}}a_e\,e^{\sqrt{\frac{V_0}{3}}\,t},
\eeq
which, as we can see from Fig.~\ref{acelt1,logacelt1}, is always positive. Close to the end of inflation we have
\beq
\label{apptapprox}
\ddot{a}(t)=\frac{16}{9}\left(-7+4\sqrt{3}\right)a_e\,V_0^{3/2}\,t+{\cal O}(t^2),
\eeq
and $\ddot{a}(t)$ vanishes at $t=0$. The scale factor as well as its logarithm are shown by Fig.~\ref{at1,logat1} while the acceleration and its log are shown in Fig.~\ref{acelt1,logacelt1}.
In both figures we normalize w.r.t. the value of the scale factor at the end of inflation $a_e$ thus, $a(t)/a_e=1$ and $\ddot{a}(t)=0$ at the end of inflation where $t=0$.
\begin{figure}[t!]
\par
\begin{center}
\includegraphics[trim = 0mm  0mm 1mm 1mm, clip, width=8.cm, height=5.cm]{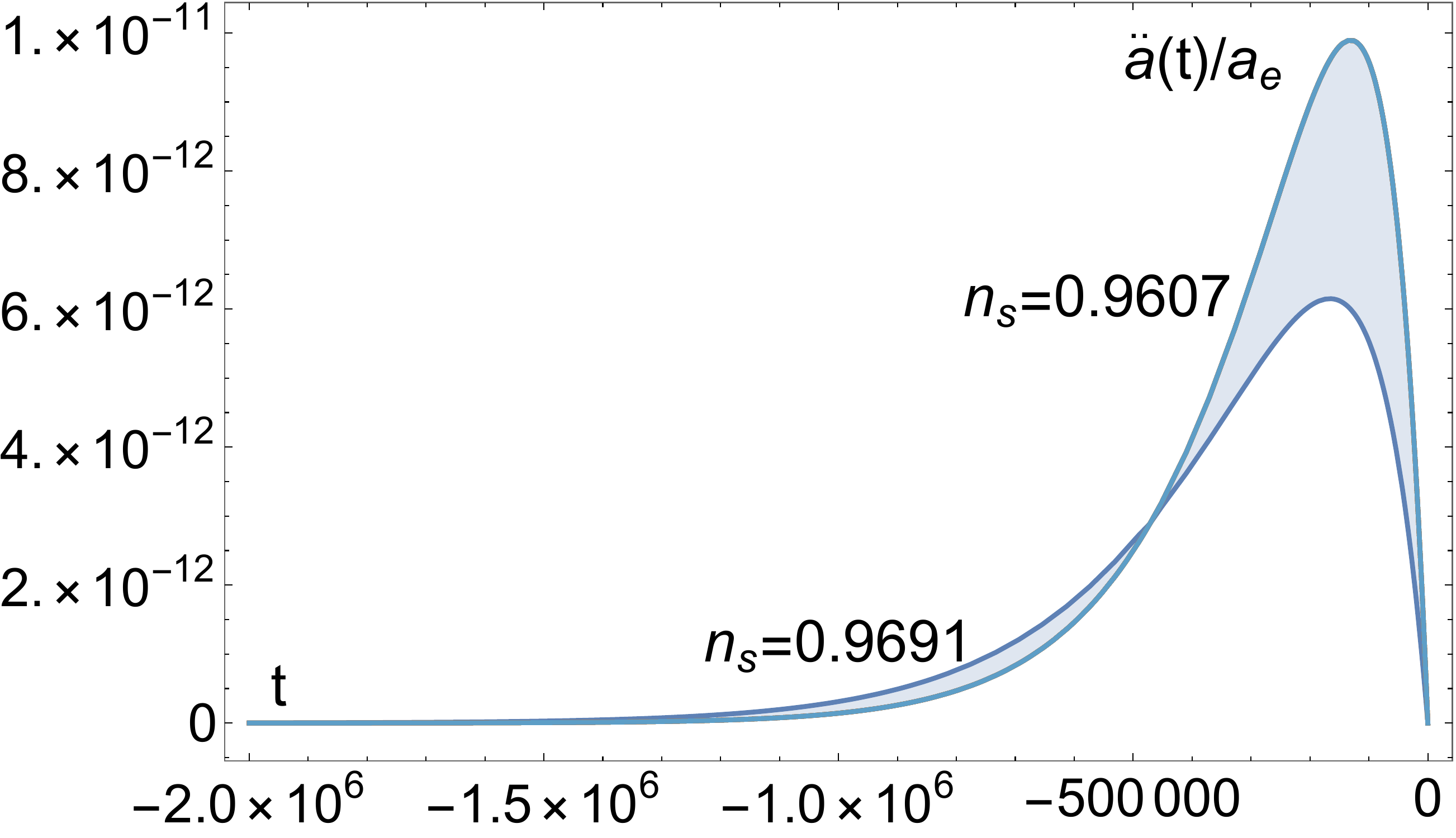}
\includegraphics[trim = 0mm  0mm 1mm 1mm, clip, width=8.cm, height=5.cm]{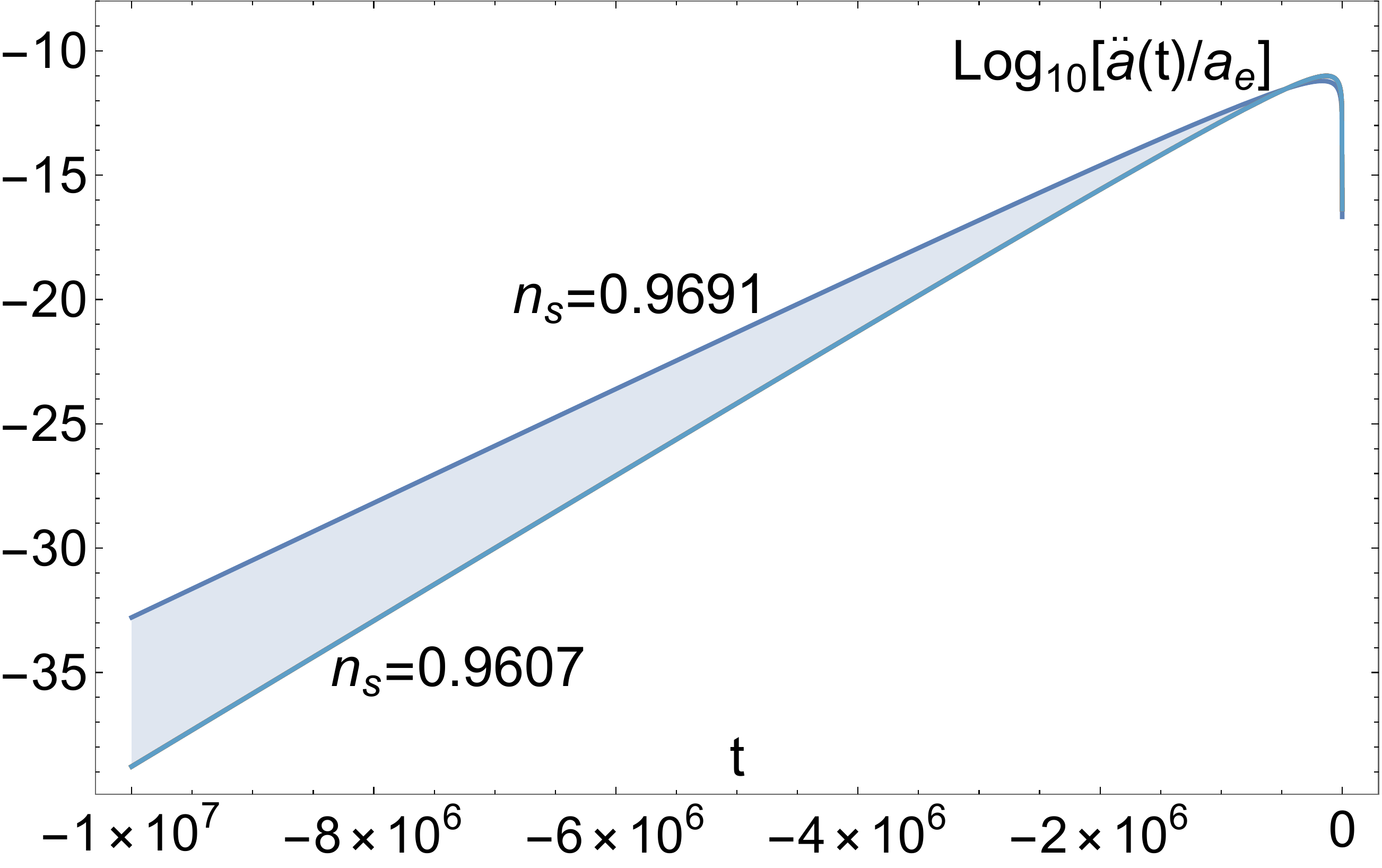}
\end{center}
\caption{The acceleration of the scale factor normalized by the scale factor at the end of inflation and its logarithm as functions of time. The quasi-exponential behavior of $\ddot{a}(t)$ is clearly visible in the logarithmic plot. Close to the end of inflation the acceleration decreases rapidly to a vanishing value for $t=0$ which signals the end of inflation. The shadowed region as described in Fig.~\ref{fit0}.
}
\label{acelt1,logacelt1}
\end{figure}
The Hubble function is
\beq
\label{Ht}
H(t)=\frac{(6-4\sqrt{3})(-3+2\sqrt{V_0}\,t)\sqrt{V_0}}{9+12(-2+\sqrt{3})\sqrt{V_0}\,t},
\eeq
and it is shown in Fig.~\ref{Ht1},
$H(t)$ decreases very slowly during most of inflation. We can also calculate the number of e-folds as a function of time from $a(t)$ up to the end of inflation at $a_e$, the result is
\beq
\label{Nt}
N(t)\equiv \ln(\frac{a_e}{a(t)})= \ln\left(\frac{e^{-\sqrt{\frac{V_0}{3}}\,t}}{\left(1+\frac{4}{3}(-2+\sqrt{3})\sqrt{V_0}\,t \right)^{3/4}}\right).
\eeq
One can show (using Eqs.~(\ref{fit}) and (\ref{fik})) that this expression for $N(t)$ is the same as $N_{ke}$ as given by Eq.~(\ref{efolds}).  We see that the biggest contribution to $N(t)$ occurs at the beginning of inflation when the exponential is large (being the time $t$ negative) decreasing near the end of inflation as
\beq
\label{Ntapprox}
N(t)\approx \left(2-\frac{4}{3}\right)\sqrt{V_0}\,t+{\cal O}(t^2),
\eeq
thus for the range of $t$ given in Eq.~(\ref{bounds}) the number of e-folds is bounded as $48.9<N(t)<62.5$. 
\subsection{The field evolution} \label{STAfi} 
\begin{figure}[tb]
\begin{center}
\includegraphics[width=8cm]{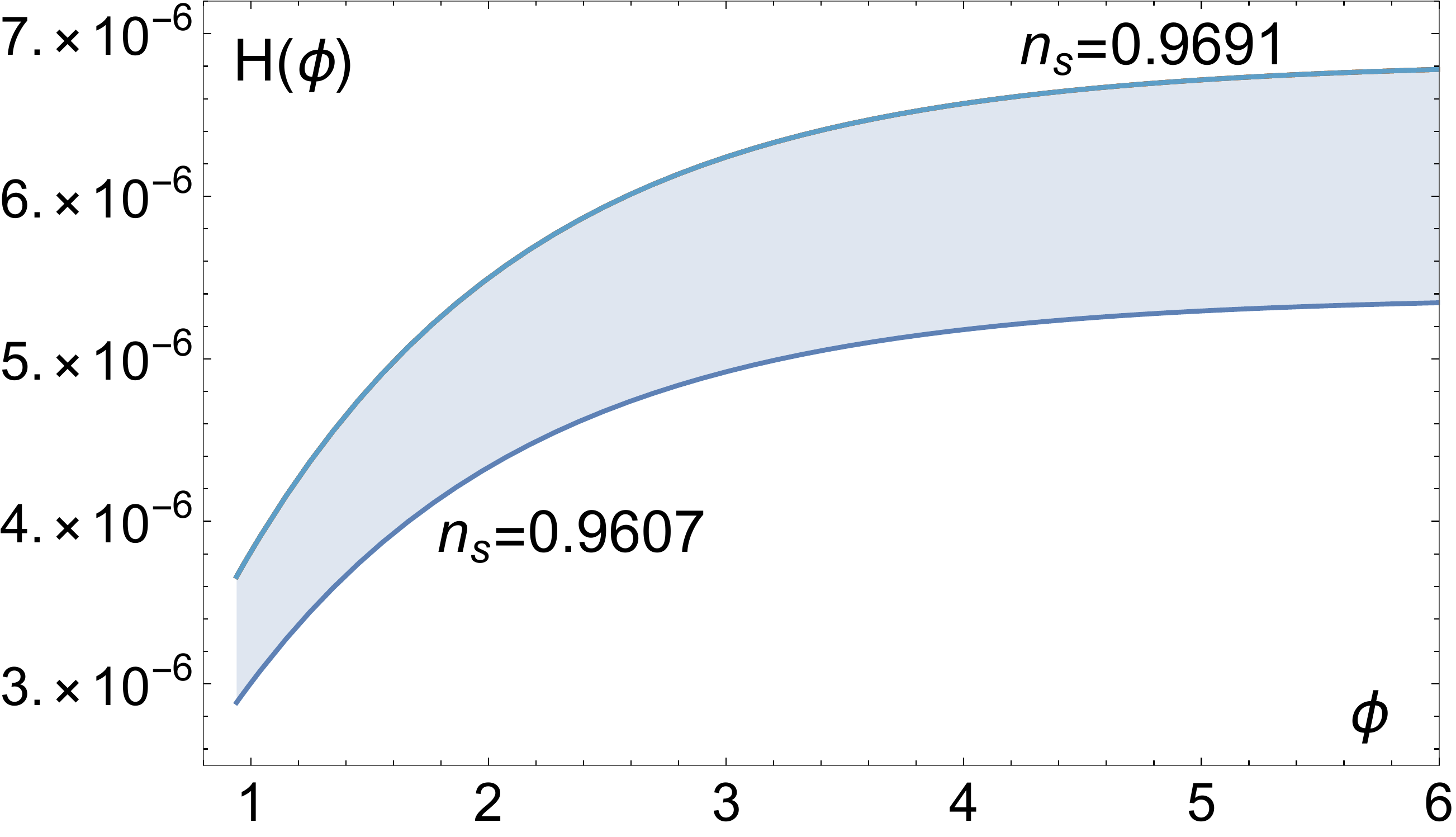}
\caption{\small Plot of the Hubble function as a function of the inflaton field $\phi$. From the time scales the size of the pivot scale leave the horizon to the end of inflation the Hubble function decreases about half its initial value. The shadowed region as described in Fig.~\ref{fit0}. }
\label{hfi1}
\end{center}
\end{figure}

We can invert Eq.~(\ref{fit}) to get
\beq
\label{tfi}
t= \frac {3}{4}\sqrt{\frac {3}{V_0}} \left(1+\frac{2}{\sqrt{3}}-e^{\sqrt{\frac {2}{3}}\phi(t)} \right) .
\eeq
Using Eq.~(\ref{at}) we get an expression for the scale factor as function of the inflaton
\beq
\label{afi}
a(\phi)=\left(2\sqrt{3}-3\right)^{3/4}a_e\,e^{\frac{1}{4}\left(3+2\sqrt{3}-3e^{\sqrt{\frac{2}{3}}\,\phi}+\sqrt{6}\,\phi\right)},
\eeq
it is easy to show that for $\phi=\phi_e$, $a(\phi)$ reduces to $a_e$. Note that $a(\phi)$ does not depend on $V_0$ like most other quantities thus, no shadowed region appears in Fig.~(\ref{afi1,logafi1}).
\begin{figure}[t!]
\par
\begin{center}
\includegraphics[trim = 0mm  0mm 1mm 1mm, clip, width=8.cm, height=5.cm]{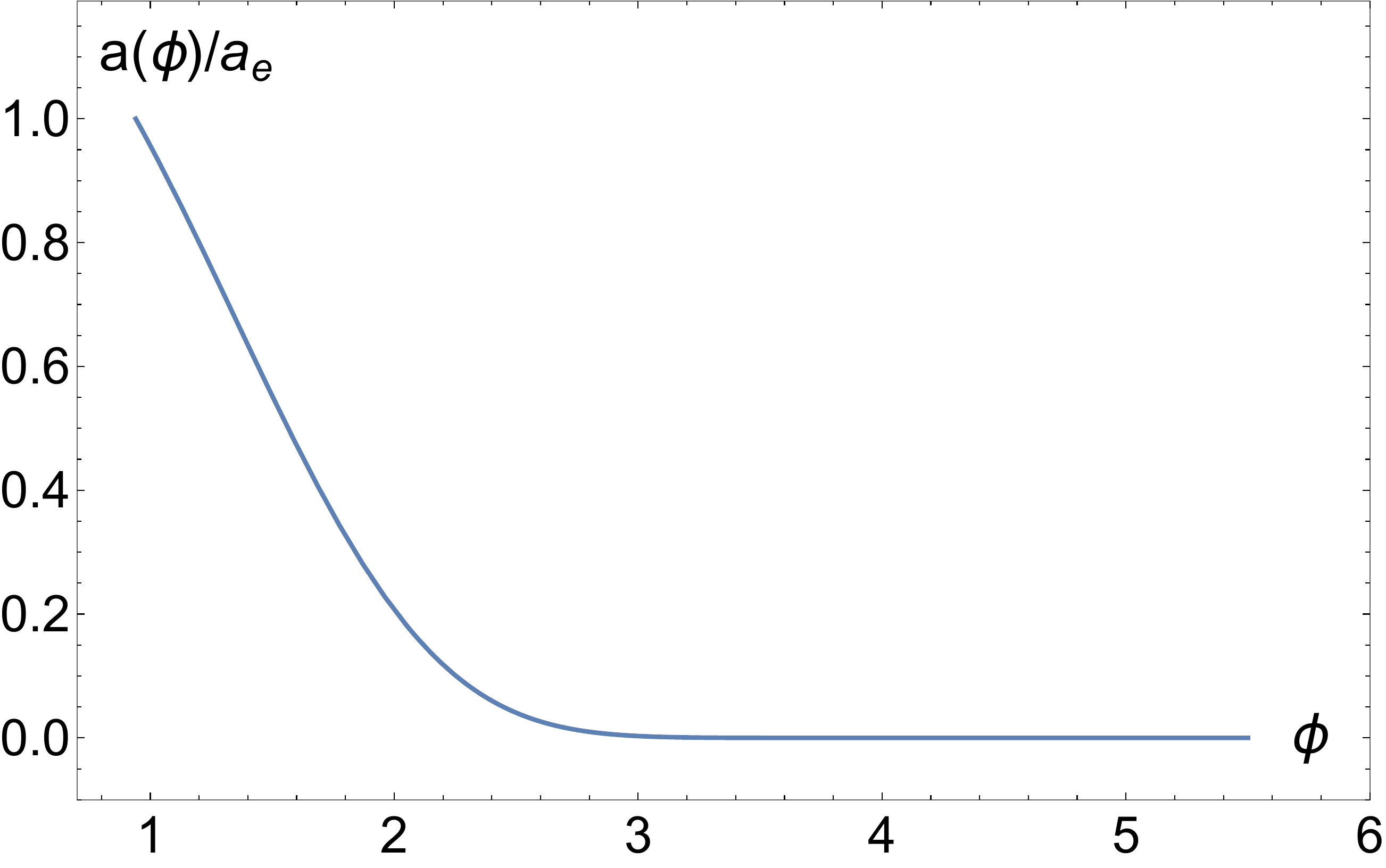}
\includegraphics[trim = 0mm  0mm 1mm 1mm, clip, width=8.cm, height=5.cm]{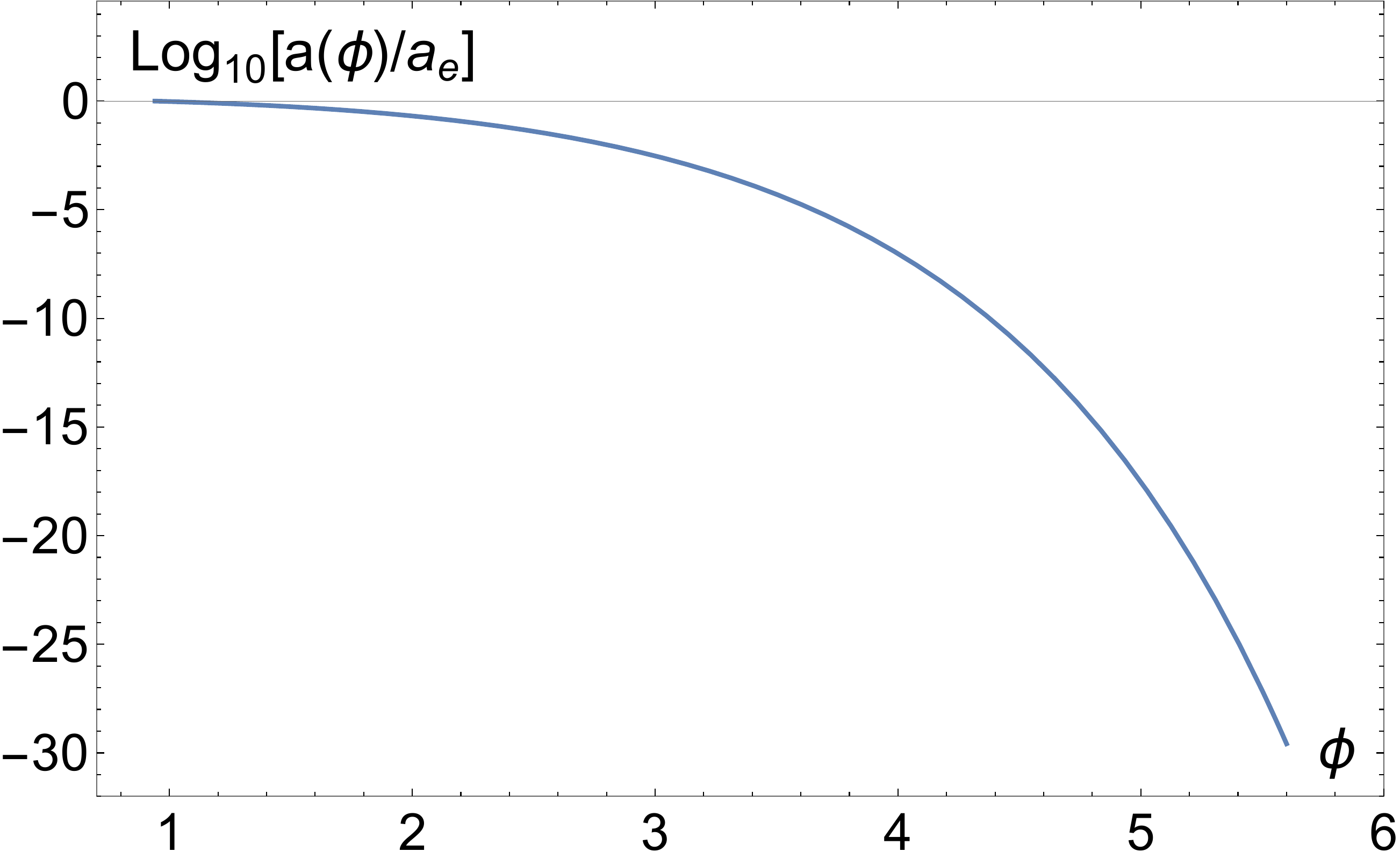}
\end{center}
\caption{The scale factor $a(\phi)$ and its logarithm normalized by the scale factor at the end of inflation $a_e$ as functions of $\phi$. The l.h.s. panel apparently shows a larger growth of $a(\phi)$ for small $\phi$ but this is not really so as is evident from the logarithmic plot in the r.h.s. panel. For example, the scale factor at $\phi=4$ is about ten orders of magnitude larger than the scale factor at  $\phi=5$ while between  $\phi=1$ and $\phi=2$ the scale factor has growth less than one order of magnitude.
}
\label{afi1,logafi1}
\end{figure}
\begin{figure}[t!]
\par
\begin{center}
\includegraphics[trim = 0mm  0mm 1mm 1mm, clip, width=8.cm, height=5.cm]{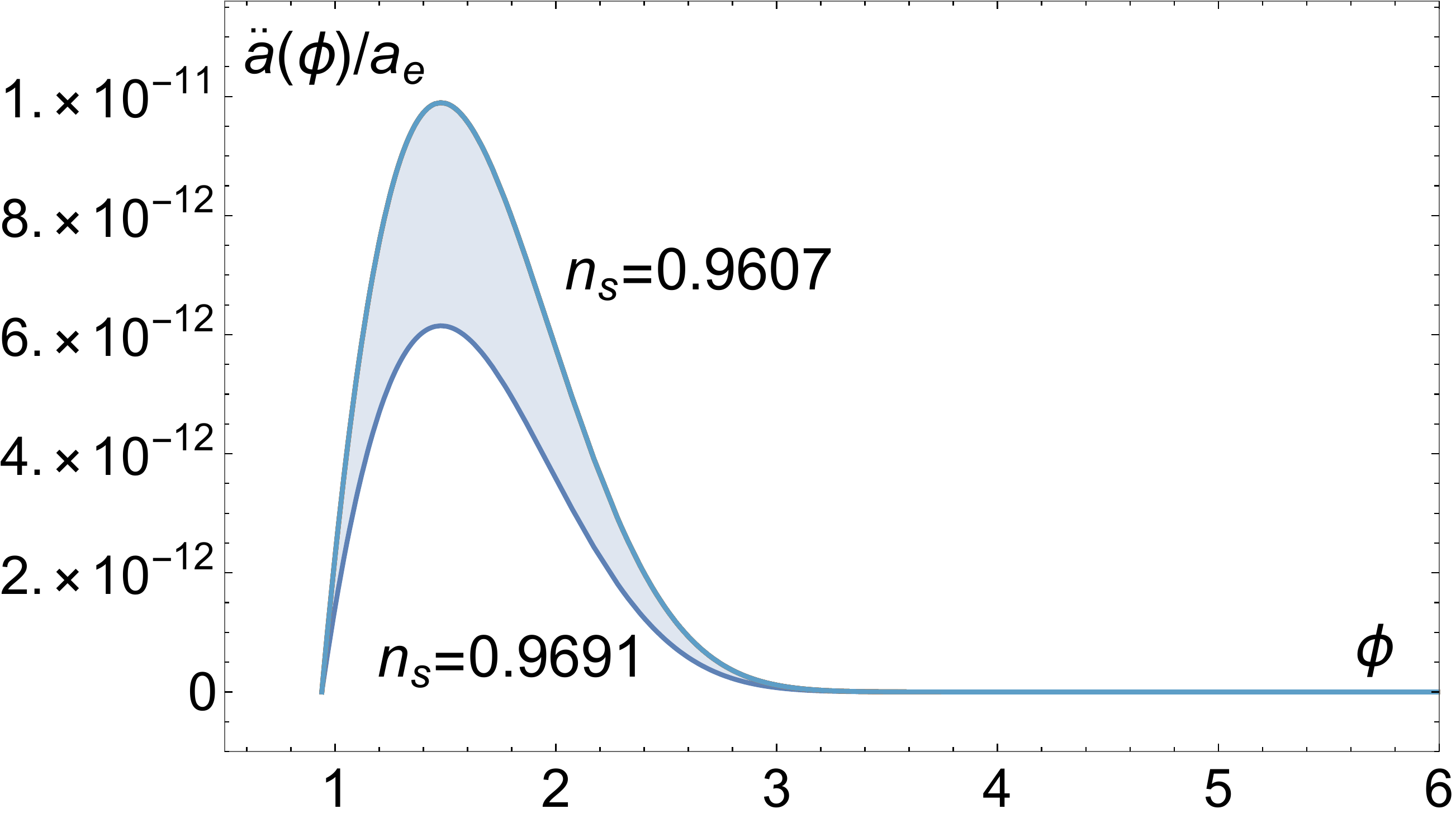}
\includegraphics[trim = 0mm  0mm 1mm 1mm, clip, width=8.cm, height=5.cm]{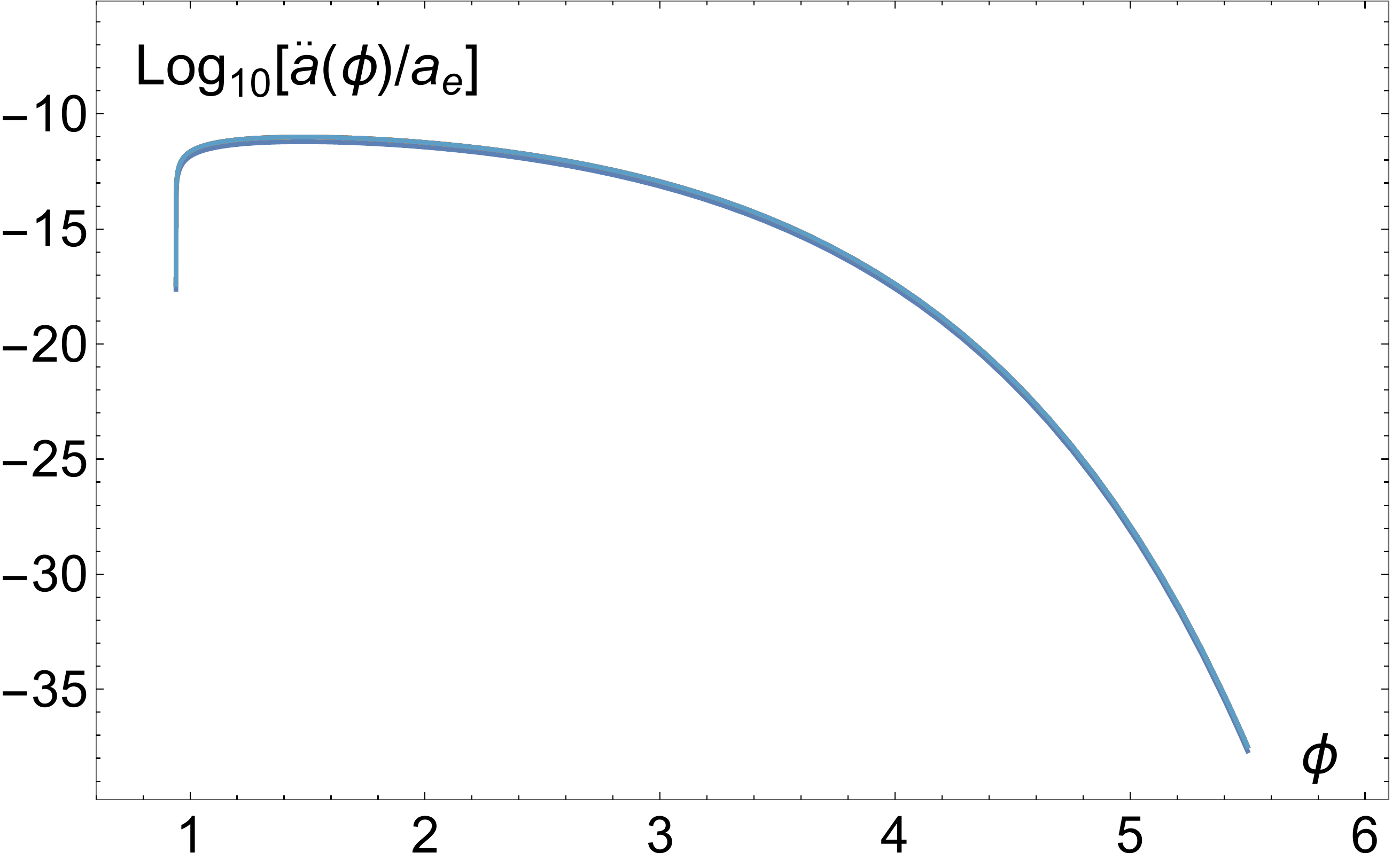}
\end{center}
\caption{We plot the acceleration of the scale factor and its logarithm normalized by the scale factor at the end of inflation $a_e$. The acceleration drops rapidly for $\phi$ close to the end of inflation at $\phi_e$. The shadowed region as described in Fig.~\ref{fit0}.
}
\label{acelfi1,logacelfi1}
\end{figure}
From Eqs.~(\ref{fit}) and (\ref{acelt}) we obtain the corresponding expression for the acceleration of the scale factor as a function of $\phi$
\beq
\label{acelfi}
\ddot{a}(\phi)=\frac{a_eV_0}{3\sqrt{3}\left(45+26\sqrt{3}\right)^{1/4}}\,e^{\frac{1}{12}\left(9+6\sqrt{3}-9\,e^{\sqrt{\frac{2}{3}}\,\phi}-5\sqrt{6}\,\phi\right)}\left(-1+3\,e^{\sqrt{\frac{2}{3}}\,\phi}\left(-2+e^{\sqrt{\frac{2}{3}}\,\phi}\right)\right)
\eeq
this is ahown in Fig.~(\ref{acelfi1,logacelfi1}) where we can see that $\ddot{a}(\phi)$ decreases fast just before the end of inflation where it vanishes. Finally, it is easy to obtain an expression for the Hubble function during inflation starting right from the original potential of Eq.~(\ref{pot}), the result is
\beq
\label{Hfi}
H(\phi)=\sqrt{\frac{V_0}{3}} \left(1-e^{-\sqrt{\frac{2}{3}}\,\phi}\right),
\eeq
Fig.~(\ref{hfi1}) shows its behavior decreasing with $\phi$ up to the value $\phi_e= \sqrt{\frac {3}{2}}\ln\left(1+\frac{2}{\sqrt{3}}\right) \approx 0.94$.
\section{Conclusions} \label{CON} 
We have written general equations for the Friedmann and fluid equations during the inflationary epoch in the slow-roll (SR) approximation in terms of the potential energy of  a scalar field. These equations are applicable to any model of inflation. From these equations solutions can be obtained for both, the scalar field and the scale factor of the universe as functions of time. As an example we have studied in detail the solutions to the Starobinsky model during inflation. Quantities such as the acceleration of the scale factor, the equation of state parameter (EoS) and the Hubble function have been given through closed analytical expressions both as a function of time and also of the scalar field $\phi$. The behavior of all these quantities has been illustrated by means of the relevant figures. 

{\em Acknowledgements:} 
We acknowledge financial support from UNAM-PAPIIT,  IN104119, {\it Estudios en gravitaci\'on y cosmolog\'ia}.

\end{document}